\documentclass[aps,prb,reprint,showpacs]{revtex4-1}
\usepackage{bm}
\usepackage{graphicx}
\usepackage{cancel}
\usepackage{braket}
\usepackage{mathrsfs}

\pdfoutput=1

\renewcommand{\vec}[1]{{\mathbf{#1}}}

\begin{document}

\title{Accurate and efficient computation of the Kohn-Sham orbital kinetic energy density in the full-potential linearized augmented plane wave method}

\author{Lin-Hui Ye}
\affiliation{Key Laboratory for the Physics and Chemistry of Nanodevices, Department of Electronics, Peking University, Beijing 100871, P.R. China}

\date{\today}

\begin{abstract}
The Kohn-Sham orbital kinetic energy density $\tau_\sigma(\vec{r}) = \sum_{i} w_{i\sigma} \big|\nabla \psi_{i\sigma}(\vec{r}) \big|^2$ is one fundamental quantity for constructing  meta-generalized gradient approximations (meta-GGA) for use by density functional theory.  We present a computational scheme of $\tau_\sigma(\vec{r})$ for full-potential linearized augmented plane wave method. Our scheme is highly accurate and efficient and easy to implement to existing computer code. To illustrate its performance, we construct the Becke-Johnson meta-GGA exchange potentials for Be, Ne, Mg, Ar, Ca, Zn, Kr, Cd atoms which are in very good agreement with the original results. For bulk solids, we construct the Tran-Blaha modified Becke-Johnson potential (mBJ) and confirm its capability to calculate band gaps, with the reported bad convergence of the mBJ potential being  substantially improved. The present computational scheme of $\tau_\sigma(\vec{r})$ should also be valuable for developing other meta-GGA's in FLAPW as well as in similar methods utilizing atom centered basis functions.
\end{abstract}


\maketitle

\section{Introduction}

Nowadays, density functional theory\cite{DFT} has become the dominant method for electronic structure calculations. Accompanying its 50 year development, the quality of the approximated energy functionals are also constantly improved.\cite{Perspective}  In order of higher accuracy, Perdew has categorized various functionals in terms of  the ``Jacob's ladder'':\cite{ladder} The first rung of the ladder is the local density approximation (LDA) which depends only on the spin densities $(\rho_\uparrow,\rho_\downarrow)$. The second rung consists of generalized gradient approximations (GGA) which depend on $(\rho_\uparrow, \rho_\downarrow, \nabla\rho_\uparrow, \nabla\rho_\downarrow)$. The third rung consists of  ``meta generalized gradient approximations (meta-GGA)'' which further include the Kohn-Sham orbital kinetic energy density $\tau_\sigma(\vec{r})$, so that the functionals now depend on  $(\rho_\uparrow, \rho_\downarrow, \nabla\rho_\uparrow, \nabla\rho_\downarrow, \tau_\uparrow, \tau_\downarrow)$.

As a fundamental component of meta-GGA, $\tau_\sigma(\vec{r})$ should be implemented in a way which is highly accurate and efficient at the same time. For the full-potential linearized augmented planewave method\cite{Slater, Anderson, NUFLAPW} a previous implementation of $\tau_\sigma(\vec{r})$ exists\cite{TBimplm}. By definition $\tau_\sigma(\vec{r})$ requires the gradients of all occupied Kohn-Sham orbital ($w$ is the occupation number):
\begin{eqnarray}
\label{deftau}
\tau_\sigma(\vec{r}) = \sum_{i=1} w_{i\sigma} \Big|\nabla \psi_{i\sigma}(\vec{r}) \Big|^2
\end{eqnarray}
Since direct computation of the gradients on dense meshes in the unit cell would be too costly, the authors make use of the Kohn-Sham equation
\begin{eqnarray}
\label{KSeq}
\left\{-\frac{\nabla^2}{2} + v_{\sigma}(\vec{r})\right\}\psi_{i\sigma}(\vec{r}) =\varepsilon_{i\sigma}\psi_{i\sigma}(\vec{r})
\end{eqnarray}
to convert $\tau_\sigma(\vec{r})$ to the following form
\begin{eqnarray}
\label{TBtau}
\tau_\sigma(\vec{r})=\frac{1}{2}\nabla^2 \rho_{\sigma}(\vec{r})  - v_{\sigma}(\vec{r})  \rho_{\sigma}(\vec{r})  -\sum_{i=1} \varepsilon_{i\sigma}\rho_{i\sigma}(\vec{r})\quad\quad
\end{eqnarray}
which involves only the total spin density $\rho_\sigma(\vec{r})$, the Kohn-Sham effective potential $v_{\sigma}(\vec{r})$ and the Kohn-Sham orbital eigenvalues $\varepsilon_{i\sigma}$.  Eq.(\ref{TBtau}) is the form being implemented. It is more efficient than Eq.(\ref{deftau}) since the dependence on individual orbital has been removed.

Nevertheless, it is often unnoticed that the conversion from  (\ref{deftau})  to (\ref{TBtau}) is not always valid, since typically not all electrons states are solved by Kohn-Sham equation. In fact, core states are usually solved with relativistic corrections which is the case in the standard FLAPW implementation. Then, expression (\ref{TBtau}) is even not non-negative-definite. Invalid numeric values are mostly found around nuclei where the Dirac-Kohn-Sham solution deviates the most from the Kohn-Sham solution. Technically, these invalid data may be zero'ed without affecting the band structure very much, for which the dominant effects come from potential in the chemical bonding regions. It is unavoidable, however, that the convergence of self-consistency is often ruined, and some author questions whether the bad convergence is intrinsic to the meta-GGA exchange potential itself.\cite{convg}

Of course, one can intentionally use Kohn-Sham equation to solve not only the valence states but also the core states. With the sacrifice of the relativistic corrections,  Eq.(\ref{TBtau}) becomes equivalent to Eq.(\ref{deftau}). Even so, however, the equivalence is purely mathematical rather than numerical. This is because in almost all cases Kohn-Sham equation is solved only approximately by use of finite basis sets. Through the conversion from (\ref{deftau}) to (\ref{TBtau}) the incomplete basis set error is brought into the values of $\tau_\sigma(\vec{r})$.

It is always desirable to calculate $\tau_\sigma(\vec{r})$ directly from the original definition Eq.(\ref{deftau}) since this equation is valid regardless of the details of how the orbital are solved. The purpose of this work is to present a computational scheme of $\tau_\sigma(\vec{r})$  for full-potential linearized augmented planewave method (FLAPW). Our scheme is highly accurate and efficient and requires minimum change to the existing computer code. Moreover, the increased accuracy of $\tau_\sigma(\vec{r})$  substantially improves the convergence of self-consistency.

This paper is organized as the following: In section II the FLAPW method is briefly reviewed. As this method has become very popular in recent years and there have been numerous literatures about it,\cite{FLAPWreview} our introduction is limited to the minimum content which is relevant to the present work. In section III, we derive the formulae of our computational scheme of $\tau_\sigma(\vec{r})$  for both the valence and the core states. The performance of this scheme is illustrated in section IV. We first construct the Becke-Johnson meta-GGA exchange potential\cite{BJ06} for atoms, and then the Tran-Blaha modified version of the potential\cite{mBJ} for bulk materials, by which we discuss its major feature for calculating semiconductor band gaps. We summarize our work in Section V. Within the FLAPW framework, the unit cell is partitioned into atomic spheres (called the muffin-tin region) and the space in between (called the interstitial region). Implementation of $\tau_\sigma(\vec{r})$ in the interstitial region is simple and straightforward. Therefore, most discussions are focused on the muffin-tin region. Throughout this work atomic units are used. To simplify the notations, spin index and some other symbols are suppressed wherever confusion is avoided.

\section{Brief Introduction to the FLAPW method}

The FLAPW method solves the valence states and core states by different techniques. Valence states are expanded by the following basis functions:
\begin{widetext}
\begin{eqnarray}
\label{basis}
\phi_{\vec{g}}(\vec{r}) = \left\{\begin{array}{ll}
\sum\limits_{lm} \bigg[ a_{l}(\vec{g})  u_l(r) + b_{l}(\vec{g}) \dot{u}_l(r) \bigg] Y_{lm}^{\star}(\hat{\vec{g}})Y_{lm}(\hat{\vec{r}})  & (\mbox{muffin-tins}) \\[2.ex]
\frac{1}{\sqrt{\Omega}}e^{i\vec{g}\cdot\vec{r}} & (\mbox{interstitial})
\end{array}
\right.
\end{eqnarray}
\end{widetext}
Each basis function is of a hybrid form: Inside the muffin-tins it takes the linear combination of atomic orbital, while within the interstitial it is simply plane wave. In this fashion, the basis function best accounts for the general feature of the wave functions in the whole space which behave like atomic wave functions close to nuclei and become free-electron-like in-between. The optimal shape of the basis function allows the expansion of the valence state
\begin{eqnarray}
\label{FLAPWexpand}
\psi_{n\vec{k}}(\vec{r}) = \sum_{\vec{G}} z_{n\vec{k}}(\vec{\vec{G}}) \phi_{\vec{g}}(\vec{r}), \hskip 1cm (\vec{g}=\vec{k}+\vec{G})
\end{eqnarray}
to use small cutoff for the reciprocal lattice vector $\vec{G}$.

In (\ref{basis}), $Y_{lm}$ is spherical harmonics  which is widely used in atom centered basis functions. The $u_l(r)$ is solved by the radial differential equation with a pre-chosen energy parameter and keeping only the spherical component of the potential. The combination  $u_l(r)Y_{lm}(\hat{\vec{r}})$ is called a muffin-tin function. The other radial function $\dot{u}_l(r)$ is the energy derivative of $u_l(r)$. It serves as the first order correction to $u_l(r)$, so that the basis function becomes applicable to a wide energy range. The use of the two muffin-tin functions inside one atomic sphere is proposed by Anderson\cite{Anderson} as the linearised approximation to the original augmented plane wave method of Slater\cite{Slater}. The linear coefficients $a_l(\vec{g})$ and $b_l(\vec{g})$ are so chosen that the basis function and its first derivative are both continuous across the sphere boundary.

To account for space group symmetry, density  and potential are expanded by symmetrized plane waves (called star functions) in the interstitial region and symmetrized spherical harmonics (called lattice harmonics) in the muffin-tin region. In particular, within an atomic sphere, a lattice harmonics is constructed as
\begin{eqnarray}
\label{lattharm}
K_{\nu}(\hat{\vec{r}})=\sum_{m_{\nu}}c_{\nu} (m_{\nu}) Y_{l_{\nu}m_{\nu}}(\hat{\vec{r}})
\end{eqnarray}
so that it is invariant under any local  point group symmetry $\mathcal{R}_i$:
\begin{eqnarray}
\label{rilatt}
\mathcal{R}_i K_{\nu}(\hat{\vec{r}}) = K_{\nu}(\hat{\vec{r}})
\end{eqnarray}
Expand the density and the potential by the lattice harmonics
\begin{eqnarray}
\label{rholattharm}
\rho(\vec{r}) &=& \sum_{\nu} \rho_{\nu}(r) K_{\nu}(\hat{\vec{r}}) \\
\label{vlattharm}
v(\vec{r}) &=&  \sum_{\nu} v_{\nu}(r) K_{\nu}(\hat{\vec{r}})
\end{eqnarray}
symmetry requirements are automatically fulfilled.

By applying $\mathcal{R}_i$ to Eq.(\ref{TBtau}), we realize that $\tau(\vec{r})$ has the same symmetry  as $\rho(\vec{r})$ and $v(\vec{r})$, and therefore can also be expanded by the same set of lattice harmonics
\begin{eqnarray}
\label{taulatharm}
\tau(\vec{r}) = \sum_{\nu}\tau_{\nu}(r)K_{\nu}(\hat{\vec{r}})
\end{eqnarray}
The expansion coefficients are calculated by:
\begin{eqnarray*}
\tau_{\nu}(r) = \int \tau(\vec{r}) K^\star_{\nu}(\hat{\vec{r}})d\Omega
\end{eqnarray*}

Core states in the standard FLAPW implementation are solved by Dirac-Kohn-Sham equation:
\begin{eqnarray}
\label{DKS}
\left\{ c\bm{\alpha}\cdot\vec{p} +c^2\beta + v^{(0)}(r) \right\}\psi_{\kappa\mu}(\vec{r}) = \varepsilon_{\kappa\mu}\psi_{\kappa\mu}(\vec{r})\quad\quad
\end{eqnarray}
to better account for relativistic effects. In (\ref{DKS}), $\bm{\alpha}$ and $\beta$ are the standard Dirac matrices.  $v^{(0)}(r)$ only contains the spherical component of $v(\vec{r})$. The four component solution to Eq.(\ref{DKS}) is
\begin{eqnarray}
\label{fourcomp}
\psi_{\kappa\mu}^c(\vec{r}) =\left(
\begin{array}{c}
g_{i\kappa}(r)\chi_{\kappa\mu}(\hat{\vec{r}}) \\[2.ex]
if_{i\kappa}(r)\hat{\sigma}_r\chi_{\kappa\mu}(\hat{\vec{r}})
\end{array}
\right)
\end{eqnarray}
with $g_{n\kappa}(r)$ and  $f_{n\kappa}(r)$ being the radial part of the major and minor components, respectively.  Note that the core states are assumed to be completely restricted within the muffin-tins so that their wave functions have no interstitial part. Besides, $g_{n\kappa}(r)$ and  $f_{n\kappa}(r)$ are not expanded by basis set, but are directly integrated on the real space grid.

The total kinetic energy density is contributed by the core states and all occupied valence states:
\begin{eqnarray}
\label{tautot}
\tau(\vec{r}) = \tau^c(\vec{r}) + \tau^v(\vec{r})
\end{eqnarray}

\section{Implementation scheme of $\tau(\vec{r})$ for the FLAPW method}
\subsection{$\tau^v(\vec{r})$ of the muffin-tin region: contribution from the valence states}
Within the muffin-tin region, the expansion of the valence state can be written as:
\begin{eqnarray}
\psi_{n\vec{k}}(\vec{r})= \sum\limits_{lm} \bigg[A_{lm}(n\vec{k})  u_l(r) + B_{lm}(n\vec{k}) \dot{u}_l(r) \bigg] Y_{lm}(\hat{\vec{r}})\quad\quad\quad\quad
\end{eqnarray}
in which we have defined the coefficients
\begin{eqnarray*}
A_{lm}(n\vec{k}) &=& \sum_{\vec{G}}z_{n\vec{k}}(\vec{g})\:a_l(\vec{g})\:Y^\star_{lm}(\hat{\vec{g}}) \\
B_{lm}(n\vec{k}) &=& \sum_{\vec{G}}z_{n\vec{k}}(\vec{g})\:b_l(\vec{g})\:Y^\star_{lm}(\hat{\vec{g}})
\end{eqnarray*}
The total kinetic energy density of the valence states inside the muffin-tin region can be written as:
\begin{widetext}
\begin{eqnarray}
\label{taunk}
\tau^v(\vec{r}) = \sum_{n\vec{k}}w_{n\vec{k}}\Big|\bm{\nabla}\psi_{n\vec{k}}(\vec{r})\Big|^2 =  \nabla^2 \rho^v(\vec{r}) -\sum_{n\vec{k}}w_{n\vec{k}}\bigg\{\psi_{n\vec{k}}^\star(\vec{r})\nabla^2\psi_{n\vec{k}}(\vec{r})  - \psi_{n\vec{k}}(\vec{r})\nabla^2\psi_{n\vec{k}}^\star(\vec{r})\bigg\}
\end{eqnarray}
\end{widetext}
Note that to reach at (\ref{taunk}) we did not make use of any physics equation such as Kohn-Sham equation. Therefore, (\ref{taunk}) is equivalent to (\ref{deftau}) regardless of the details of how the orbital are solved.

On the right hand side of Eq.(\ref{taunk}), the first term is already calculated by GGA. The remaining two terms can be derived by use of the following relation:
\begin{widetext}
\begin{eqnarray}
\label{favorprop}
\nabla^2 \Big\{W(r)Y_{lm}(\hat{\vec{r}})\Big\} = \left( W''(r) + \frac{2W'(r)}{r} - \frac{l(l+1)W(r)}{r^2} \right)Y_{lm}(\hat{\vec{r}})
\end{eqnarray}
\end{widetext}
Eq.(\ref{favorprop}) is a favorable property of all muffin-tin functions, by which the laplacian operation only affects the radial part while leaves the angular part unaltered. Using (\ref{favorprop}) for (\ref{taunk}), we  get $\tau^v(\vec{r})$ in the muffin-tin region. Projected to the lattice harmonics representation, the final expression of the expansion coefficients is:
\begin{widetext}
\begin{eqnarray}
\label{tauv}
\tau^v_{\nu}(r) &=&  \sum_{ll'}\Bigg\{ \bigg[\;u'_{l}u'_{l'} + \frac{l(l+1)+l'(l'+1)-l_\nu(l_\nu+1)}{2r^2}u_{l}u_{l'}\bigg]\sum_{n\vec{k}}w_{n\vec{k}}\sum_{mm'}A_{l'm'}^\star A_{lm}\Big\langle Y_{lm}\Big|Y_{l_\nu m_\nu}\Big|Y_{l'm'}\Big\rangle \nonumber\\
&&\hskip 0.6cm +  \bigg[\:\dot{u}'_{l}\dot{u}'_{l'} + \frac{l(l+1)+l'(l'+1)-l_\nu(l_\nu+1)}{2r^2}\dot{u}_{l}\dot{u}_{l'}\bigg]\sum_{n\vec{k}}w_{n\vec{k}}\sum_{mm'}B_{l'm'}^\star B_{lm}\Big\langle Y_{lm}\Big|Y_{l_\nu m_\nu}\Big|Y_{l'm'}\Big\rangle   \nonumber\\
&&\hskip 0.6cm +  \bigg[\:u'_{l}\dot{u}'_{l'} + \frac{l(l+1) + l'(l'+1)-l_\nu(l_\nu+1)}{2r^2}u_{l}\dot{u}_{l'} \bigg]\sum_{n\vec{k}}w_{n\vec{k}}\sum_{mm'}A_{l'm'}^\star B_{lm}\Big\langle Y_{lm}\Big|Y_{l_\nu m_\nu}\Big|Y_{l'm'}\Big\rangle \nonumber\\
&&\hskip 0.6cm +  \bigg[\:\dot{u}'_{l}u'_{l'} + \frac{l(l+1) + l'(l'+1)-l_\nu(l_\nu+1)}{2r^2}\dot{u}_{l}u_{l'}\bigg]\sum_{n\vec{k}}w_{n\vec{k}}\sum_{mm'}B_{l'm'}^\star A_{lm}\Big\langle Y_{lm}\Big|Y_{l_\nu m_\nu}\Big|Y_{l'm'}\Big\rangle \Bigg\}
\end{eqnarray}
For comparison, we also write out the expansion coefficients of the density:
\begin{eqnarray}
\label{rhov}
\rho^v_{\nu}(r) &=&  \sum_{ll'}\Bigg\{ \bigg[\;u_{l}u_{l'} \bigg]\sum_{n\vec{k}}w_{n\vec{k}}\sum_{mm'}A_{l'm'}^\star A_{lm}\Big\langle Y_{lm}\Big|Y_{l_\nu m_\nu}\Big|Y_{l'm'}\Big\rangle \nonumber\\
&&\hskip 0.6cm +  \bigg[\:\dot{u}_{l}\dot{u}_{l'}\bigg]\sum_{n\vec{k}}w_{n\vec{k}}\sum_{mm'}B_{l'm'}^\star B_{lm}\Big\langle Y_{lm}\Big|Y_{l_\nu m_\nu}\Big|Y_{l'm'}\Big\rangle   \nonumber\\
&&\hskip 0.6cm +  \bigg[\:u_{l}\dot{u}_{l'} \bigg]\sum_{n\vec{k}}w_{n\vec{k}}\sum_{mm'}A_{l'm'}^\star B_{lm}\Big\langle Y_{lm}\Big|Y_{l_\nu m_\nu}\Big|Y_{l'm'}\Big\rangle \nonumber\\
&&\hskip 0.6cm +  \bigg[\:\dot{u}_{l}u_{l'} \bigg]\sum_{n\vec{k}}w_{n\vec{k}}\sum_{mm'}B_{l'm'}^\star A_{lm}\Big\langle Y_{lm}\Big|Y_{l_\nu m_\nu}\Big|Y_{l'm'}\Big\rangle \Bigg\}
\end{eqnarray}
\end{widetext}
Eqs.(\ref{tauv}) and (\ref{rhov}) are similar. Their differences are limited to the radial parts while their angular parts are the same. This is not a coincidence, but a consequence of the property (\ref{favorprop}) of the muffin-tin functions. Eq.(\ref{tauv}) is the expression being implemented in the present work. It is highly accurate because it is equivalent to Eq.(\ref{deftau}) and no extra error is introduced throughout the derivation. Moreover, the expressoin of the lattice harmonics expansion also ensures the correct symmetry of $\tau(\vec{r})$.

Eq.(\ref{tauv}) is equally efficient because $\tau^v_{\nu}(r)$ can be calculated together with $\rho^v_{\nu}(r)$. During the construction of the valence density, almost all computational cost of (\ref{rhov}) is spent on the angular part, while the radial part takes no more than a few percent. By calculating $\tau_{\nu}^{v}(r)$ together with $\rho^v_{\nu}(r)$, the costly angular part needs to be evaluated only once. The extra cost for the radial part of (\ref{tauv}) is very small. Note that the radial part of (\ref{tauv}) has no orbital dependence since it only requires the radial basis functions $u_l(r)$ and $\dot{u}_l(r)$ which are universal for all valence states. Technically, the implementation of the radial part of (\ref{tauv}) to the construction of the valence density is also straightforward and requires little human labor.

\subsection{$\tau^c(\vec{r})$ of the muffin-tin region: contribution from the core states}

Core states also contribute to  the $\tau(\vec{r})$ of the muffin-tin region.  Although core states are in the four-component form, each row is a separate muffin-tin function so that Eq.(\ref{favorprop}) is equally applicable. At ground state, all core sub-shells are fully filled. Consequently, core density is spherical and $\tau(\vec{r})$ also is because they must have the same angular parts:
\begin{eqnarray}
\tau^c(\vec{r}) = \tau_0^c(r)
\end{eqnarray}
The expansion coefficient of the first lattice harmonics $K_0(\hat{\vec{r}})$ is contributed by all occupied sub-shells:
\begin{eqnarray}
\label{tauc}
\tau_{0}^c(r) =\sum_{l} \tau^c_{0,l}(r)
\end{eqnarray}
\begin{widetext}
For the $l=\kappa$ sub-shell:
\begin{eqnarray}
\label{tauikappa1}
\tau^c_{0,l}(r)=\frac{2l}{\sqrt{4\pi}}\Bigg\{\:{g'_{i\kappa} }^2 + \frac{l(l+1)}{r^2} g_{i\kappa}^2 + {f'_{i\kappa} }^2 + \frac{l(l-1)}{r^2} f_{i\kappa}^2 \Bigg\}
\end{eqnarray}
For the $l=-\kappa-1$ sub-shell:
\begin{eqnarray}
\label{tauikappa2}
\tau^c_{0,l}(r)=\frac{2(l+1)}{\sqrt{4\pi}}\Bigg\{{g'_{i\kappa} }^2 + \frac{l(l+1)}{r^2} g_{i\kappa}^2  + {f'_{i\kappa} }^2 + \frac{(l+1)(l+2)}{r^2} f_{i\kappa}^2\Bigg\}
\end{eqnarray}
\end{widetext}
Since the radial solution of Dirac-Kohn-Sham equation is slightly irregular, $\tau(\vec{r})$ diverges at the nucleus which is similar to that of all GGA potentials. Alternatively, core states can also be solved by Kohn-Sham equation. Then the expression of $\tau_0^c(r)$ can be directly derived from the valence expression (\ref{tauv}) in which one only needs to set $l_{\nu}=0$ and $l=l'$ and remove the linearization term $\dot{u}_l(r)$.

\subsection{$\tau^v(\vec{r})$ of the interstitial region}

In the interstitial region, only valence states contribute to $\tau(\vec{r})$:
\begin{widetext}
\begin{eqnarray}
\label{tauint}
\tau^v(\vec{r})= \frac{1}{\Omega}\sum\limits_{n\vec{k}} w_{n\vec{k}}  \sum\limits_{\vec{G}_i\vec{G}_j} \Big(\vec{k}+\vec{G}_i\Big)\Big(\vec{k}+\vec{G}_j\Big) z_{n\vec{k}}^\star(\vec{G}_i)z_{n\vec{k}}(\vec{G}_j) \: e^{i(\vec{G}_j-\vec{G}_i)\cdot\vec{r}}
\end{eqnarray}
\end{widetext}

Eqs.(\ref{tauv}), (\ref{tauikappa1}), (\ref{tauikappa2}) and (\ref{tauint}) have been implemented to the FLAPW code of Northwestern University.\cite{NUFLAPW} To test the performance of our computational scheme, we next construct the  Becke-Johnson type meta-GGA potentials for atoms and bulk semiconductors. For atoms, core states are solved by Kohn-Sham equation for comparison with the earlier work. For semiconductors, both Kohn-Sham equation and Dirac-Kohn-Sham equation are tested for solving the core states, and we find that the results as well as the speed of convergence are nearly the same. Therefore, throughout this work all core states are solved by Kohn-Sham equation.

\section{The Becke-Johnson type meta-GGA potentials}

\subsection{The atomic potentials}
The Becke-Johnson (BJ06) potential\cite{BJ06} is a meta-GGA exchange potential attempting to approach the optimized effective potential (OEP)\cite{OEP} of atoms. BJ06 consists of two terms:
\begin{eqnarray}
\label{BJ06}
v^{\mathrm{BJ06}}_{x,\sigma}(\vec{r}) = v^{\mathrm{BR89}}_{x,\sigma}(\vec{r}) + \frac{1}{\pi}\sqrt{\frac{12}{5}}\sqrt{\frac{\tau_\sigma(\vec{r})}{\rho_\sigma(\vec{r})}}
\end{eqnarray}
The first term is the Becke-Roussel exchange potential\cite{BR89} which is derived by assuming hydrogen-like exchange hole. $v^{\mathrm{BJ06}}_{x,\sigma}(\vec{r})$ is determined by a nonlinear equation involving $\rho_{\sigma}(\vec{r}), \nabla\rho_{\sigma}(\vec{r}), \nabla^2\rho_{\sigma}(\vec{r})$ and $\tau_\sigma(\vec{r})$, and closely resembles the Slater averaged exchange.\cite{simpleHF} Compared to OEP, $v^{\mathrm{BR89}}_{x,\sigma}(\vec{r})$ is too deep and lacks the characteristic shell-structures. Therefore, the second term is added for correction. This term  is repulsive which reduces the strength of the exchange. Besides, it  strongly varies across atomic shells which restores the OEP shell structures. Summing up the two terms, the BJ06 potential is close to the accuracy of the OEP of atoms, yet is much simpler since it is directly constructed from local quantities rather than being solved by the complicated OEP equation. Nevertheless, BJ06 is a model exchange potential without corresponding exchange energy. Therefore, it cannot be used for calculating total energy.

\begin{figure}[htp]
\centering
\includegraphics[width=\linewidth]{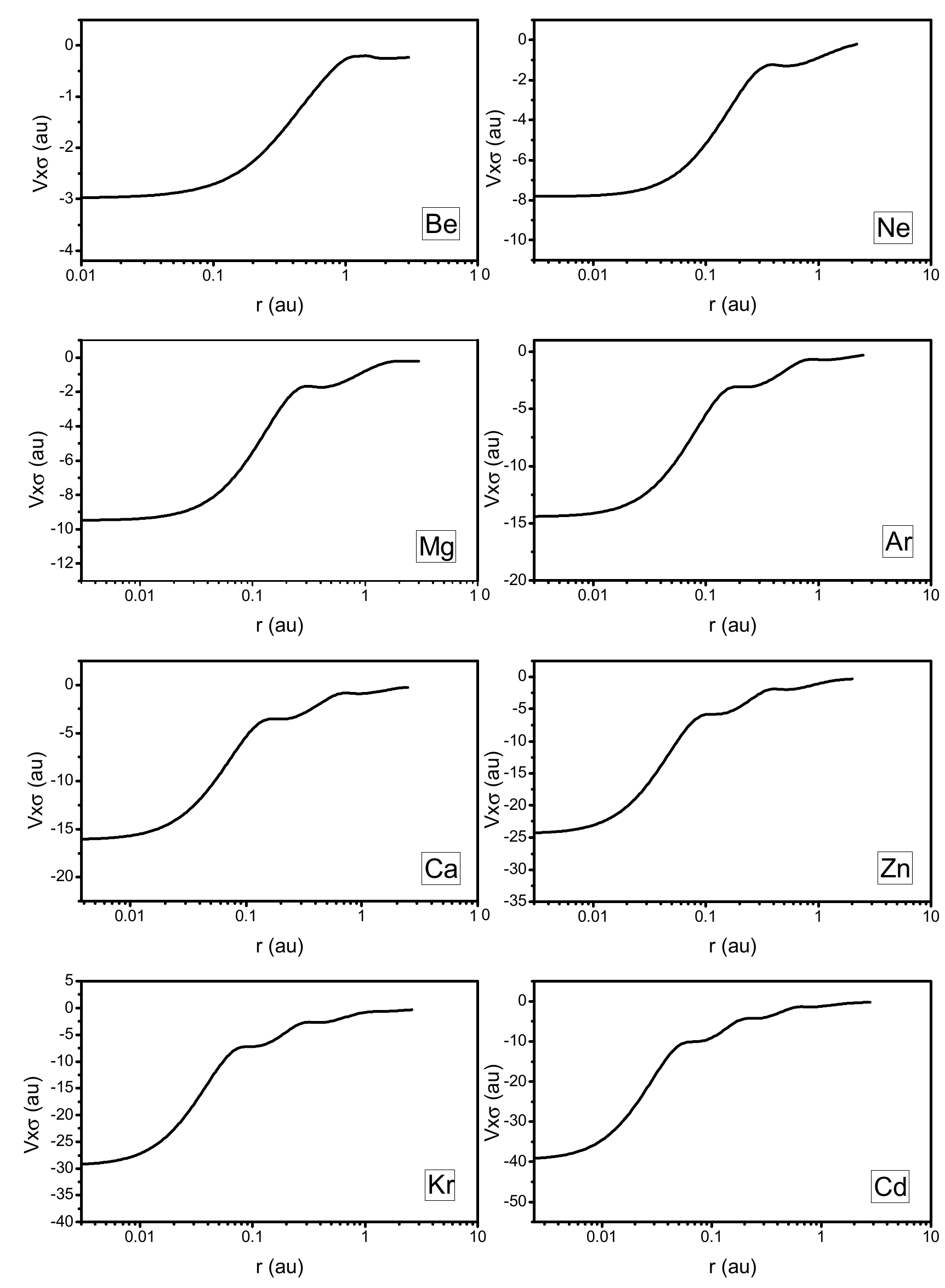}
\caption{Becke-Johnson exchange potential for Be, Ne, Mg, Ar, Ca, Zn, Kr, Cd atoms calculated by the FLAPW method with the implementation of $\tau_\sigma(\vec{r})$ of this work.\label{atombj06}}
\end{figure}

For Be, Ne, Mg, Ar, Ca, Zn, Kr, Cd atoms, the BJ06 potentials have been published by Becke and Johnson.\cite{BJ06} We have generated the same potentials by our FLAPW code and the results are plotted in Figure \ref{atombj06}.  As can be seen, both the location and  the height of the ``bumps'' between atomic shells are well reproduced.  Besides, close to the nuclei all potentials have the correct, asymptotically flat shape, indicating the adequacy of the treatment of $\tau(\vec{r})$ in the muffin-tin region. The depth of the potential bottom also agrees well with the original results. Overall, all atomic potential are in excellent agreement with Ref.\cite{BJ06}. The remaining slight differences may be ascribed to the use of different boundary conditions in the two groups of calculations: While the original calculations are performed with the molecular code, the present calculations use the FLAPW code with supercells and periodic boundary condition.

\subsection{The bulk potentials}
To test our implementation of  $\tau(\vec{r})$ for bulk materials, we construct the Tran-Blaha modified version of the BJ06 potential (mBJ)\cite{mBJ} for solids. The TB09 exchange is essentially the same as BJ06, with only the weights of the BR89 term and the shell term being changed:
\begin{eqnarray}
\label{TB09}
v_{x,\sigma}^{\mathrm{TB09}}(\vec{r}) = cv_{x\sigma}^{\mathrm{BR89}}(\vec{r})
+ (3c-2)\frac{1}{\pi}\sqrt{\frac{5}{12}}\sqrt{\frac{\tau_\sigma(\vec{r})}{\rho_\sigma(\vec{r})}}
\end{eqnarray}
In Eq.(\ref{TB09}) the weights  are determined by the $c$ parameter, which is essentially adjustable, but can also be calculated ``pseudo-{\it ab initio}-ly'' by empirical formulae proposed by the authors\cite{ImproveMBJ}. For $c=1$  TB09  reverts to the original BJ06.  The total mBJ potential is formed either by TB09 exchange alone, or by combining it with a correlation potential such as LDA. It is such slight adjustment of the weights which leads to the surprising discovery that the mBJ potential is capable of calculating semiconductor band gaps. Albeit being a local potential so that its computational cost is essentially maintained at the LDA level, in many cases the mBJ potential can achieve band gap accuracy which is even comparable to the  much more sophisticated GW approximation.

Applications of the mBJ potential has been quickly growing in recent years which also motivates the present work. However, the existing FLAPW implementation uses  Eq.(\ref{TBtau})  to calculate  $\tau(\vec{r})$, and therefore suffers from the problems mentioned in the Introduction. It is thus desirable to recheck the mBJ potentials by our implementation of  $\tau(\vec{r})$. In all our calculations, the $c$ parameter of (\ref{TB09}) is automatically determined by the suggested formula:\cite{ImproveMBJ}
\begin{eqnarray}
\label{cformular}
c= A + B\bar{g}
\end{eqnarray}
with $A=0.488, \quad B=0.5$, and
\begin{eqnarray}
\label{gbar}
\bar{g}=\frac{1}{V_{\mathrm{cell}}}\int\limits_{\mathrm{cell}} \frac{1}{2}\left(\frac{|\nabla \rho^\uparrow(\vec{r})|}{\rho^\uparrow(\vec{r})}+\frac{|\nabla \rho^\downarrow(\vec{r})|}{\rho^\downarrow(\vec{r})}\right)d^3\vec{r}
\end{eqnarray}
For the correlation potential we use LDA.\cite{VWN}  One difference with the earlier work is that in our calculations spin-orbit coupling (SOC) is included perturbatively through the second variational approach. Typically, the effect of SOC on the value of band gap is smaller than 0.1 eV. But for materials (such as ZnTe) with heavy elements SOC can change the band gap by as much as 0.3 eV.
\begin{figure}
\centering
\includegraphics{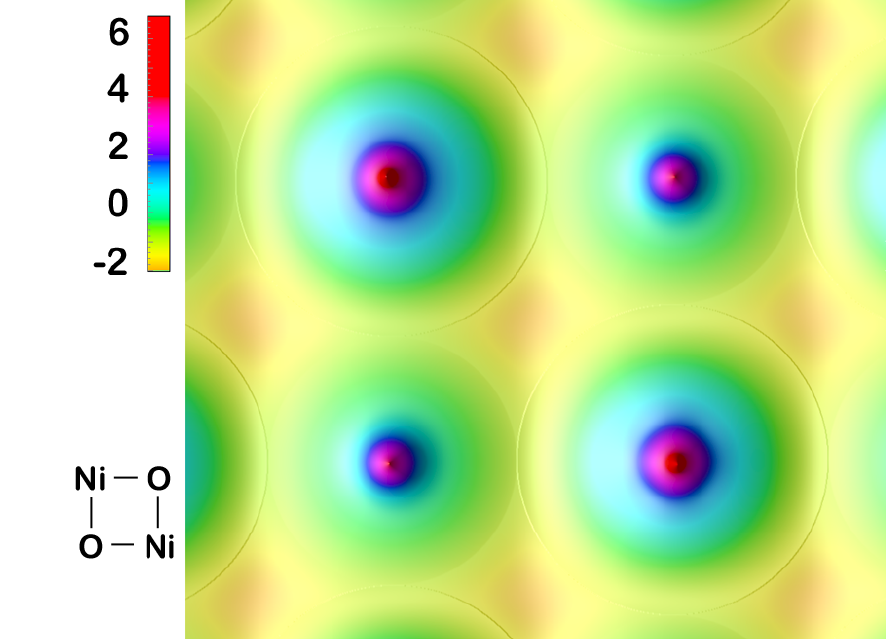}
\caption{(Color online) Rubber-sheet plot of the up-spin kinetic energy density $\log_{10}\tau_{\uparrow}$ of the NiO (001) plane.  Atomic positions are illustrated at the left bottom of the plot.\label{tau}}
\end{figure}

\begin{figure}
\includegraphics{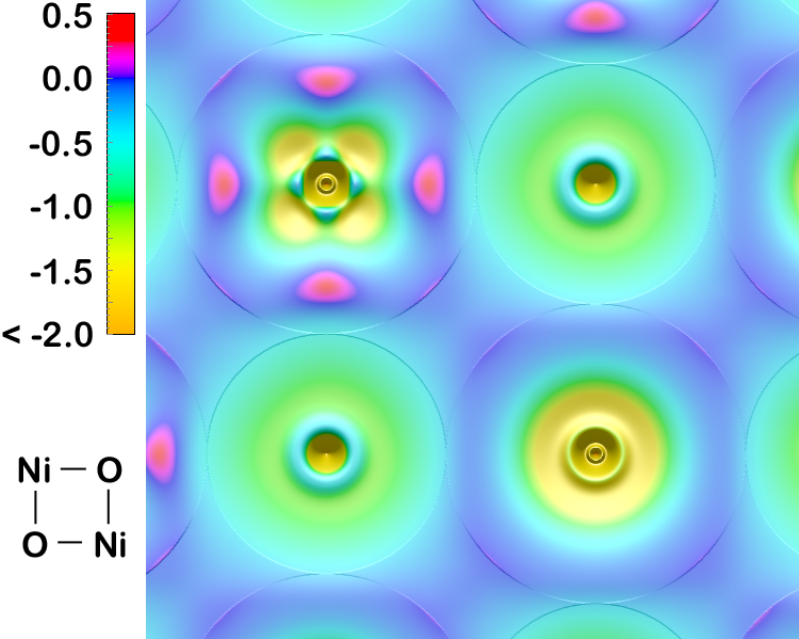}
\caption{(Color online) Rubber-sheet plot of the up-spin mBJ potential $v_{\uparrow}^{\mathrm{mBJ}}$ of the NiO (001) plane.  Atomic positions are illustrated at the left bottom of the plot.\label{NiO}}
\end{figure}

In Figures \ref{tau} and \ref{NiO} we plot\cite{OpenDX} the spin-up kinetic energy density and the mBJ potential of the NiO (001) plane for comparison with the similar plot of Ref.\cite{MeritsMBJ}. At the nuclei $\tau(\vec{r})$  shows sharp spikes while $v^{\mathrm{mBJ}}(\vec{r})$ achieves the minimum, although neither quantities diverges. The present implementation of $\tau(\vec{r})$ allows us to reveal even the finest details of the mBJ potential in Figure \ref{NiO}. Especially,  all shell structures are clearly seen. Another feature is that, although $\tau(\vec{r})$ and $v^{\mathrm{mBJ}}(\vec{r})$ are very smooth within both the muffin-tin and the interstitial regions, across the sphere boundary they are obviously discontinuous. Such discontinuity is also observed in the potential plot of NiO in Ref.\cite{MeritsMBJ} and is irrelevant to the potential itself or to the implementation of $\tau(\vec{r})$. Rather, it is caused by the discontinuity of the basis functions. By (\ref{basis}), the FLAPW basis function is continuous only to the first derivative while the meta-GGA require the second derivative of the density. In fact, the basis function is already discontinuous even at the zero'th order because in Eq.(\ref{basis}) the summation is cutoff at finite $(lm)$. By the same reason, the GGA potential\cite{PBE} is also discontinuous across sphere boundary, although we notice that usually the discontinuity of the meta-GGA is more serious than the GGA.
\begin{figure}[htp]
\centering
\includegraphics[width=8cm]{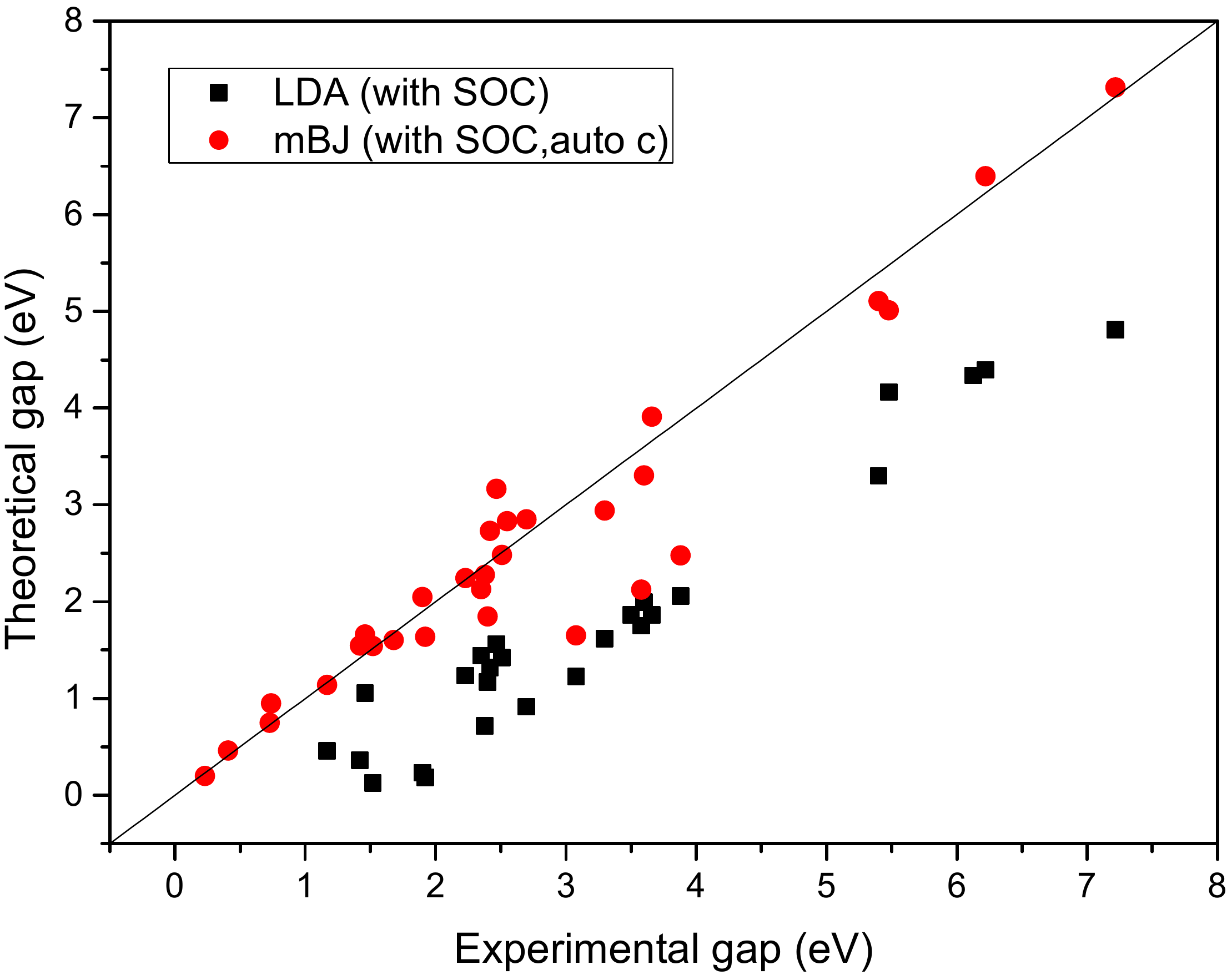}
\caption{(Color online) Comparison between the mBJ band gaps, the LDA band gaps and the experimental band gaps for the 40 semiconductors of the SC40 test set.\label{sc40}}
\end{figure}

With our highly accurate $\tau(\vec{r})$, we have calculated the mBJ band gaps for all semiconductors in the SC40 test set.\cite{sc40} The results are plotted in Figure \ref{sc40} together with the LDA band gaps to compare with experiment. Indeed, for all semiconductors mBJ systematically improves the band gaps. Especially, the improvement is nearly as good for the smallest band gap (of InSb with $E_g=0.23$ eV) as for the largest (of MgO with $E_g=7.31$ eV). Besides, we also confirm that in most cases the size for the band gap increases monotonically with increasing $c$. Nevertheless, serious errors still exist for the three Ba compounds,  BaS, BaSe, BaTe, suggesting that there might be problems in the Ba potential or the experimental data.

We have thus found that the improvement of $\tau(\vec{r})$ does not qualitatively change the most important feature of the mBJ potential. This is not surprising since the improvement of $\tau(\vec{r})$ is most significant around the nuclei while the band gap correction mainly depends on potential far away.\cite{MeritsMBJ}  In more details, the dominant contribution to the band gap correction comes from the shell term which is repulsive so that it pushes all band states upward. Away from the nuclei the shell term  increases and for open systems it approaches a positive constant at $r\rightarrow\infty$. Because the conduction states are usually more delocalized than the valence states, the upshift of the conduction states by the shell term is larger than the valence states, and therefore the band gap is enlarged. Through this mechanism, band gap correction is not substantially affected by the improvement of $\tau(\vec{r})$ around the nuclei. For the same reason, it is neither affected very much by the divergence of $\tau(\vec{r})$ if core states are solved by Dirac-Kohn-Sham equation.

It has been reported before that self-consistency by the mBJ potential is harder to achieve than LDA or GGA.\cite{convg} This is confirmed in our work. Typically, using the mBJ potential 50\% more iterations are needed to achieve self-consistency than LDA or GGA.  For an example, for MoS$_2$ LDA takes 22 iterations to converge, while mBJ with our implementation of $\tau(\vec{r})$ requires 32 iterations. Nevertheless, we have also tried implementing   $\tau(\vec{r})$ by Eq.(\ref{TBtau}), by which we found that the speed of convergence is drastically ruined. For MoS$_2$ the mBJ calculation with Eq.(\ref{TBtau}) requires 79 iterations to converge, while calculation of NiO does not even converge at all. In fact, in all our testings, the implementation of  $\tau(\vec{r})$ by (\ref{deftau}) always converges better than by (\ref{TBtau}), and the speed of convergence is not sensitive to whether the core states are solved by Kohn-Sham equation or Dirac-Kohn-Sham equation. Besides, for more than 70 solids we have calculated with (\ref{deftau}), all converge well except for CoO and FeO which fail because mBJ is incapable of lifting the near degeneracy of the $3d$ states. Therefore, it is safe to conclude that at least a large part of the reported convergence problem is not intrinsic to the potential but is caused by technical reasons.

\section{Summary}

In this work we have presented an highly accurate and efficient computational scheme for the Kohn-Sham orbital kinetic energy density $\tau_\sigma(\vec{r})$ to the full-potential linearised augmented plane wave method. To test its performance, we have constructed the Becke-Johnson exchange potential for atoms which are in very good agreement with the original results. For bulk solids we have constructed the Tran-Blaha modified Becke-Johnson potential for semiconductors and confirmed its capability to calculate semiconductor band gaps. As to the convergence problem  reported before, we found that a large part  is due to the impropriate implementation of $\tau(\vec{r})$.  With respect to accuracy, efficiency, easiness of implementation and speed of convergence, our scheme all supersedes the existing implementation.  We expect this work to be valuable for developing other meta-GGA's in FLAPW as well as in similar methods utilizing atom centered basis functions.

\section{Acknowledgement}

This work is supported by the Ministry of Science and Technology of China (Grant Nos. 2011CB933001 and 2011CB933002).

\end{document}